\documentclass[%
 reprint,
superscriptaddress,
 amsmath,amssymb,
 aps,
]{revtex4-2}

\usepackage{graphicx}
\usepackage{dcolumn}
\usepackage{bm}
\usepackage{titlesec}
\usepackage{xcolor}
\usepackage{lipsum}

\usepackage{todonotes}
\setuptodonotes{inline} 

\graphicspath{ {./images/} }

\begin{document}

\preprint{APS/123-QED}

\title{First Principles Study of the Electronic Structure of the Ni$_2$MnIn/InAs and Ti$_2$MnIn/InSb  interfaces}

\author{Brett Heischmidt}
\affiliation{School of Physics and Astronomy, University of Minnesota, Minneapolis, MN 55455, USA}
\author{Maituo Yu}%
\affiliation{%
 Department of Materials Science and Engineering, Carnegie Mellon University, Pittsburgh, PA 15213, USA
}%
\author{Derek Dardzinski}%
\affiliation{%
 Department of Materials Science and Engineering, Carnegie Mellon University, Pittsburgh, PA 15213, USA
}%
\author{James Etheridge}
\affiliation{School of Physics and Astronomy, University of Minnesota, Minneapolis, MN 55455, USA}
\author{Saeed Moayedpour}%
\affiliation{%
 Department of Chemistry, Carnegie Mellon University, Pittsburgh, PA 15213, USA
}%
\author{Vlad S. Pribiag}
\affiliation{School of Physics and Astronomy, University of Minnesota, Minneapolis, MN 55455, USA}
\author{Paul A. Crowell}
\affiliation{School of Physics and Astronomy, University of Minnesota, Minneapolis, MN 55455, USA}
\author{Noa Marom}
 \email{Corresponding author: nmarom@andrew.cmu.edu}
\affiliation{%
 Department of Materials Science and Engineering, Carnegie Mellon University, Pittsburgh, PA 15213, USA
}%
\affiliation{%
 Department of Chemistry, Carnegie Mellon University, Pittsburgh, PA 15213, USA
}%
\affiliation{Department of Physics, Carnegie Mellon University, Pittsburgh, PA 15213, USA}

\date{\today}

\begin{abstract}

We present a first-principles study of the electronic and magnetic properties of epitaxial interfaces between the Heusler compounds Ti$_2$MnIn and Ni$_2$MnIn and the III-V semiconductors, InSb and InAs, respectively. We use density functional theory (DFT) with a machine-learned Hubbard $U$ correction determined by Bayesian optimization. We evaluate these interfaces for prospective applications in Majorana-based quantum computing and spintronics. In both interfaces, states from the Heusler penetrate into the gap of the semiconductor, decaying within a few atomic layers. The magnetic interactions at the interface are weak and local in space and energy. Magnetic moments of less than 0.1 $\mu_B$ are induced in the two atomic layers closest to the interface. The induced spin polarization around the Fermi level of the semiconductor also decays within a few atomic layers. The decisive factor for the induced spin polarization around the Fermi level of the semiconductor is the spin polarization around the Fermi level in the Heusler, rather than the overall magnetic moment. As a result, the ferrimagnetic narrow-gap semiconductor Ti$_2$MnIn induces a more significant spin polarization in the InSb than the ferromagnetic metal Ni$_2$MnIn induces in the InAs. This is explained by the position of the transition metal $d$ states in the Heusler with respect to the Fermi level. Based on our results, these interfaces are unlikely to be useful for Majorana devices but could be of interest for spintronics.

\end{abstract}

\maketitle

\section{\label{sec:intro}Introduction}

Majorana zero modes (MZMs) have been at the forefront of device physics for over a decade due to the tantalizing possibility of using them to achieve topological, fault tolerant quantum computing \cite{Nayak2008,Akhmerov2010,Stern2013,Sarma2015,Aasen2016}.  Thus far,  the most pursued scheme  for MZM realization has been based on a semiconducting nanowire with strong spin-orbit coupling (SOC) interfaced with a superconductor \cite{Sau2010,Alicea2010,Lutchyn2010,Oreg2010}. Obtaining Majoranas in semiconductor nanowires requires a helical gap, which arises from the combined effect of the Rashba SOC and a sufficient Zeeman interaction \cite{Frolov2020,2017_Kammhuber_NatComm}. In this system, the spin channels of the quasi-1D channel's first mode are shifted due to Rashba SOC. Then, the degeneracy near $k=0$ is lifted by Zeeman splitting, induced by an external magnetic field.  This scheme further requires superconducting pairing with a chemical potential inside this Zeeman gap, giving rise to a spinless p-wave superconducting state \cite{Read2000,Kitaev2001}. After careful fine tuning, the (necessary but not sufficient) MZM signature of zero bias conductance peaks has been observed in such superconductor-semiconductor systems \cite{Mourik2012,Das2012,Nadj-perge2014}. Recently, these results have come under scrutiny with multiple alternative explanations \cite{Moore2018,Sauls2018,Chen2019,Prada2020,YuFrolov2021,Sarma2021}, in particular, stemming from system disorder.  This has motivated  several studies of these materials systems and their interfaces. 
 
Besides the challenge of mitigating defects and disorder, from a technological standpoint, utilizing a global external magnetic field further constrains device geometry due to its necessary orientation orthogonal with the Rashba field. A recent avenue of research thus attempts to replace the external magnetic field with local proximity-induced magnetic interactions  \cite{Vaitiekenas2021}. If successful, this could potentially remove restrictions on device structure and scaling. The main criterion is that the induced Zeeman splitting in the semiconductor should be on the order of 10 $\mu$eV or larger, in order to generate a substantial helical gap. A second, but equally important criterion is that the induced magnetic interactions should be present across the diameter of the wire, not just at the interface with the ferromagnet because an interfacial effect alone would not open a helical gap in the bulk of the wire and would hence be inadequate for Majoranas. Third, the ferromagnet should be insulating, in order to avoid shunting the current flowing through the Majorana device and in order to avoid screening the electrostatic gate acting on the nanowire.Work in this direction has focused primarily on the ferromagnetic insulator EuS interfaced with InAs and Al \cite{Takiguchi2019,Vaitiekenas2021}. The observation of zero bias peaks in this system has sparked an active debate regarding their origin \cite{Maiani2021,Langbehn2021,Liu2021,Khindanov2021,Escribano2021}.
In particular, experimental \cite{Liu2020} and computational \cite{YuMarom2021a} studies of the EuS/InAs interface have suggested that the proximity-induced magnetic moment in InAs may be too small and too localized to produce sufficiently strong exchange coupling. Exploration of alternatives to EuS may advance the understanding of these proposed mechanisms as they relate to the atomistic structure and electronic properties of the interface and ultimately lead to the discovery of new materials platforms for the realization of MZMs.

We have chosen to investigate Heusler compounds as alternative magnetic materials to EuS. The Heuslers boast a range of intriguing properties. For example, the Heusler family includes topological superconductors \cite{Nakajima2015}, Weyl materials \cite{Manna2018}, and skyrmionic materials \cite{Nagaosa2013,Nayak2017,Fert2017}. Heuslers with Weyl points have a large anomalous Hall angle \cite{Burkov2014,Manna2018} and negative magnetoresistance under alignment of electric and magnetic fields, known as the chiral anomaly \cite{Nielsen1983,Son2013,Burkov2015,Huang2015,Xiong2015}. The sheer number of compounds within the Heusler family \cite{Graf2011} has prompted several computational high-throughput screening efforts of bulk materials \cite{He2018,Balluff2017,Rotjanapittayakul2018,Sanvito2017,Gao2019}, based on density functional theory (DFT) with computationally efficient semi-local exchange-correlation functionals. 

The Heusler family of materials has also long attracted interest in the spintronics community, thanks to a high Fermi level spin polarization in certain cases, including some half-metallic compounds \cite{DeGroot1983,Hanssen1990,Ristoiu2000,Farshchi2013}. In the context of spin injection, a main problem for epitaxially grown ferromagnetic materials on substrates is misfit dislocations due to lattice mismatch, which reduces the activation volume \cite{FelserHeuslerBook}. Mitigation of this effect is an ongoing effort.  Further, although there has been some exploration of using non-ferromagnetic materials for spintronic applications \cite{Jungwirth2016}, this research avenue is still nascent, particularly in context of spin injection utilizing Heusler compounds \cite{Akiho2013,Palmstrom2016Review,Rath2018CMSGaAs,Peterson2016,Hoffman2019}.  This motivates further study of interfaces between semiconductors and lattice matched Heusler compounds, especially those with high Fermi-level spin polarization.  It has been demonstrated, for example, that Heusler compounds such as Co$_2$FeSi and Co$_2$MnSi show higher spin-injection efficiency than Fe \cite{Peterson2016}.  Moreover, spin transport properties of epitaxial interfaces show marked energy dependence near zero bias \cite{Chantis2007,Crooker2009,Salis2011,Fujita2019},  suggesting that interface states can have a significant impact on the performance of spintronic devices.  In extreme cases, the sign of the injected polarization can reverse upon changes in interface conditions induced by annealing \cite{Schultz2009FeGaAs}.  These observations motivate a study of Heusler compounds combined with an effort to model realistic interfaces with semiconductors.

In this work, we focus on Heuslers likely to form epitaxial interfaces with the semiconductors most utilized in the Majorana community, i.e. InAs and InSb with their relatively heavy elements and thus large SOC.  Ni$_2$MnIn and Ti$_2$MnIn are lattice matched to InAs and InSb to within 0.1\% and 0.2\%, respectively. Ni$_2$MnIn is a ferromagnetic metal with a moderate magnetic moment of 4.40 $\mu$B \cite{Webster1969}. A DFT study using the generalized gradient approximation (GGA) of Perdew, Burke, and Ernzerhof (PBE)\cite{Perdew1996,Perdew1997} has reported a Fermi level spin polarization of 32\% \cite{Qawasmeh2012}). The interface of Ni$_2$MnIn with InAs(001) has been studied experimentally.  The relatively disordered B2 state of Ni$_2$MnIn forms at low temperatures. L2$_1$ ordering (Cu$_2$MnAl-type structure) and increased Curie temperature can be achieved by increasing the growth temperature or a subsequent anneal, but at the cost of interface mixing. Point contact Andreev reflectometry (PCAR) further showed a spin polarization around 17\% in a 155 nm Ni$_2$MnIn layer, although with a relatively rough surface and possibly B2 ordering \cite{Xie2001,Xie2005,Scholtyssek2008a,Zolotaryov2009}.  PCAR measurements on Ni$_2$MnIn grown on InAs(110) showed spin polarization of 28\% \cite{Bocklage2007,Scholtyssek2007}. Ref. \cite{Kilian2000} used DFT with the local density approximation  (LDA) to the exchange-correlation functional to calculate the band structure of bulk Ni$_2$MnIn. Band matching considerations were then used to predict the spin transport at the interface with InAs, assuming Fermi level pinning slightly above the InAs conduction band minimum.  A free-electron model, described below, gives transmittance (T) along three high symmetry directions as T$_{[100]}$=0.75(0.19), T$_{[110]}$=0.82(0.19), and T$_{[111]}$=0.99(0.39) for the minority (majority) spin bands. Subsequently, Ref. \cite{Kilian2001} conducted a DFT LDA study of the (100) interface using a periodic supercell consisting of eight atomic layers of Ni$_2$MnIn and two layers of InAs.  They reported a magnetic moment of 0.01-0.03 $\mu_B$ induced in the InAs, and a possible spin injection channel near the $\Gamma$ point. 

To our knowledge, no experimental work has been done on Ti$_2$MnIn to date. We note that it is possible that Ti$_2$MnIn has not been studied experimentally because it is metastable. However, metastable phases can be grown by epitaxial templating on lattice matched substrates (\textit{e.g.,} \cite{silva2010substrate, zhang2012combinitorial, gorbenko2002epitaxial}). The properties of Ti$_2$MnIn have been studied computationally, using DFT with the PBE functional \cite{Jia2014,Ma2018,Noky2018}.  Ti$_2$MnIn has been predicted to be more stable in the CuHg$_2$Ti-type structure than in the Cu$_2$MnAl-type structure of Ni$_2$MnIn \cite{Jia2014}. In terms of its magnetic properties, Ti$_2$MnIn has been predicted to be a half metal with a gap of 0.4 eV in one spin channel  and 0.04 eV in the other spin channel, and ferrimagnetic with a small overall magnetic moment of 0.4 $\mu_B$ \cite{Ma2018}.  Ref. \cite{Ma2018} and \cite{Noky2018} reported similar magnetic moments, whereas those from Ref. \cite{Jia2014} are slightly larger. Notably, neither Ref. \cite{Jia2014} nor Ref. \cite{Ma2018} considered SOC.  Ref. \cite{Noky2018} found that SOC breaks the degeneracy of the Weyl points. Importantly, no work exists as of yet for the Ti$_2$MnIn/InSb interface.

In the following, we present a DFT analysis of the electronic structure and magnetic properties of Ni$_2$MnIn and Ti$_2$MnIn, as well as their interfaces with InAs and InSb, respectively. The local and semi-local functionals, which have been used in several studies of Heuslers, cited above, have limited accuracy. Due to the self-interaction error (SIE) \cite{perdew1981density, mori2006many, mori2008localization}, the spurious repulsion of an electron from itself, (semi-)local functionals severely underestimate the band gap, to the extent that some semiconductors are erroneously predicted to be metallic \cite{2019_Borlido_JCTC,bennett2019systematic,cohen2008fractional}. In addition, the highly localized $d$-states are destabilized by the SIE, pushing the $d$-bands to higher energies, which may affect their hybridization with other bands in compounds \cite{jiang2010first}. As a result of these limitations, (semi-)local functionals have failed to correctly predict the half-metallic character and magnetic moments for some Heuslers \cite{kandpal2006correlation, kandpal2007calculated, gao2013large}. Hybrid functionals contain a fraction of exact (Fock) exchange, which mitigates the effect of the SIE.  Hybrid functionals have been shown to provide a better description of the electronic structure of Heuslers \cite{nourmohammadi2010first, fiedler2016ternary}. However, the relatively high computational cost of hybrid functionals impedes their application for simulations of large interface models with several hundred atoms. The DFT+U approach provides a good compromise between accuracy and efficiency by adding a Hubbard U correction to a (semi-)local functional for the orbitals most affected by SIE \cite{himmetoglu2014hubbard}. Here, we use DFT+U with $U$ values determined by Bayesian optimization (BO) with an objective function designed to reproduce as closely as possible the band structures obtained from a more accurate hybrid functional \cite{YuMarom2020}.  

We find that adding the Hubbard $U$ correction improves upon the performance of semi-local DFT and produces results closer to a hybrid functional. Importantly, the computational cost of DFT+U(BO) enables simulating large interface models. For both the Ni$_2$MnIn/InAs and Ti$_2$MnIn/InSb interfaces, states from the Heusler penetrate into the gap of the semiconductor, decaying within a few atomic layers. The magnetic interactions at the interface are localized in space and energy. Weak magnetic moments of less than 0.1 $\mu_B$ are induced in the two atomic layers of the semiconductor closest to the interface. Based on this, we consider these interfaces unlikely to be useful for Majorana physics, at least via an exchange mechanism. The induced spin polarization around the Fermi level of the semiconductor depends mainly on the spin polarization around the Fermi level of the Heusler, rather than the overall magnetic moment. The ferrimagnetic narrow-gap semiconductor Ti$_2$MnIn is more spin polarized around the Fermi level than the ferromagnetic metal Ni$_2$MnIn, owing to the position of the transition metal $d$ states. Consequently, the Ti$_2$MnIn induces stronger spin polarization around the Fermi level of the InSb than the Ni$_2$MnIn induces in the InAs. This has implications for the potential for spin transport through the interface. In particular, in the Ti$_2$MnIn/InSb interface it may be possible to achieve spin switching by applying a bias, which makes it of potential relevance for spintronics.

\section{\label{sec:methods}Methods}

DFT calculations were performed using the Vienna Ab initio Simulation Package (VASP) \cite{Kresse1993} with the projector augmented wave method (PAW) \cite{Blochl1994,Kresse1999}. The generalized gradient approximation (GGA) of Perdew, Burke, and Ernzerhof (PBE) \cite{Perdew1996,Perdew1997} was employed to describe the exchange-correlation interactions among electrons with a Hubbard $U$ correction \cite{Dudarev1998} determined by Bayesian optimization (BO).\cite{YuMarom2020}  We note that although the DFT+U method is commonly used to mitigate the self-interaction error for localized $d$ and $f$ electrons, it has been found that in some cases, it is necessary to apply a Hubbard $U$ correction to $s$ and $p$ states, in particular when there is a strong hybridization \cite{campo2010extended, himmetoglu2014hubbard, stroppa2011revisiting, kulik2010systematic}. We have found that for InAs and InSb, which the PBE functional erroneously predicts to be metallic, applying a negative Hubbard $U$ correction to the $p$ orbitals opens a gap \cite{YuMarom2020}. The $U$ values of -0.5 eV for In-$p$ and -7.5 eV for As-$p$ for InAs and -0.2 eV for In-$p$ and -6.1 eV for InSb-$p$ , obtained in Ref. \cite{YuMarom2020} were used here (see also Table S1 in the SI). 

For Ni$_2$MnIn and Ti$_2$MnIn the optimal $U$ values were found as detailed below.  PBE calculations with and without $U$ for both bulk materials  used a kinetic energy cutoff value of 350 eV and a k-point mesh of $20\times20\times20$. The Heyd, Scuseria, and Enzerhof hybrid functional (HSE) \cite{Heyd2003,Heyd2006} was used to calculate the bulk band structures of Ni$_2$MnIn and Ti$_2$MnIn to serve as a reference for BO. For these calculations, the kinetic energy cutoff was set to 350 eV for Ni$_2$MnIn and 400 eV for Ti$_2$MnIn. The primitive cell Brillouin zone was sampled using the Monkhorst-Pack scheme with a $9\times9\times9$ k-point mesh for Ni$_2$MnIn and an $8\times8\times8$ k-point mesh for Ti$_2$MnIn.

For PBE+U(BO) calculations, the $U$ value applied to the $p$ orbitals of In was kept the same as for InAs for Ni$_2$MnIn and the same as for InSb for Ti$_2$MnIn. Two-dimensional BO was used to determine the $U$ values applied to the $d$ orbitals of Ni, Ti, and Mn, as described in Ref. \cite{YuMarom2020}. The hyperparameters of our BO implementation are the coefficients $\alpha_1$ and $\alpha_2$, which assign different weights to the band gap vs. the band structure in the objective function, the number of valence and conduction bands used for the calculation of the objective function, $N_b$, and the parameter $\kappa$, which controls the balance between exploration and exploitation in the upper confidence bound acquisition function.  Because Ni$_2$MnIn is a metal, only the band shape was considered in the optimization, \textit{i.e.} $\alpha_1$ was set to zero. $\kappa$ was set to 10. A range of $N_b$ values was tested with the best results obtained for $N_b$=4 and $N_b$=8, which gave similar $U$ values. $N_b$=8 was chosen because it  provides sampling both close to the Fermi level and into the $d$ manifold. This produced values of $U_{eff}^{Ni,d}$ = 7.24 eV and $U_{eff}^{Mn,d}$ = 2.14 eV.  For Ti$_2$MnIn $\alpha_1$ = 0.90 and $\alpha_2$ = 0.10 were used to balance the contributions of the band gap and band shape to the objective function. $N_b$ was set to 5 and $\kappa$ was set to 7.5. This produced values of $U_{eff}^{Ti,d}$ = 1.04 eV and $U_{eff}^{Mn,d}$ = 1.19 eV. Notably, any $U$ values greater than 2 produced no band gap, thus constraining our parameter space.  We further note a strong dependence of the $U$ values on the BO hyperparameters, as discussed in Ref. \cite{Dardzinski2022}.  $U_{eff}$ values used and Bayesian optimization convergence plots are shown in Table S1 and Figure S1, respectively.

All interface calculations employed the PBE+U(BO) approach, utilizing k-point meshes of $12\times12\times1$  for Ni$_2$MnIn/InAs  and $9\times9\times1$ for Ti$_2$MnIn/InSb.  The kinetic energy cutoff was 450 eV for Ni$_2$MnIn/InAs and 350 eV for Ti$_2$MnIn/InSb. For interface calculations, dipole corrections were employed \cite{Neugebauer1992} and symmetry simplifications were turned off. A specific set of mixing parameters  were found to be necessary for the convergence of the Ni$_2$MnIn/InAs interface: IMIX=1, AMIX=0.02, BMIX=1.0, AMIX-MAG=0.2, and BMIX-MAG=2.0. Structural relaxation was performed for three layers of the Heusler and six layers of the semiconductor until the Hellman-Feynman forces acting on ions were below 1e-03  eV/{\AA}. To describe the van der Waals interactions at the interface, the Tkatchenko-Scheffler (TS) pairwise dispersion method \cite{tkatchenko2009accurate} was used in the structural relaxation. SOC \cite{Steiner2016} was applied throughout with the $z$ spin quantization axis, perpendicular to the interface. Spin-polarized density of states and band structure plots were produced by projecting the contributions of the majority and minority spin channels onto the SOC basis vectors.

In the following, we provide a brief description of the free electron spin injection analysis introduced in Ref. \cite{Kilian2000}.  The model is based on a free electron in one dimension encountering a finite step potential, and continuing past the potential freely but with a decreased momentum.  Reflection and transmission coefficients are given by R=$(k_1-k_2)^2/(k_1+k_2)^2$ and T=1-R, where the subscripts 1 and 2 indicate the two sides of the step potential. Ref. \cite{Kilian2000} applies this to the interface of Ni$_2$MnIn/InAs, with the step potential occurring as an itinerant electron passes from Ni$_2$MnIn to InAs.  Because the bands in Ni$_2$MnIn and InAs are approximately parabolic near the Fermi level, the group velocity at the Fermi level goes like the crystal momentum.  The Fermi level in InAs is pinned slightly above the conduction band minimum in agreement with the experimental observation when InAs is interfaced with Au \cite{Brillson1986}.  The momentum values when the bands cross the Fermi level are used as the momenta on either side of the step potential.  The corresponding crystal momenta in Ni$_2$MnIn and InAs give $k_1$ and $k_2$, respectively.  This model assumes an interface orientation commensurate with the transmission direction, and a perfectly normal incidence.   A benefit of this analysis is in its simplicity. However, because this analysis is based entirely on the bulk band structure of both materials, it does not consider the effect of the atomistic arrangement of the interface on the electronic structure, including band bending, metal-induced gap states, interface states, and proximity-induced magnetism.

\section{\label{sec:results}Results and Discussion}

\subsection{\label{sec:bulk}Bulk Ni$_2$MnIn and Ti$_2$MnIn}

\begin{figure}[h]
    \centering
    \includegraphics[width=0.475\textwidth]{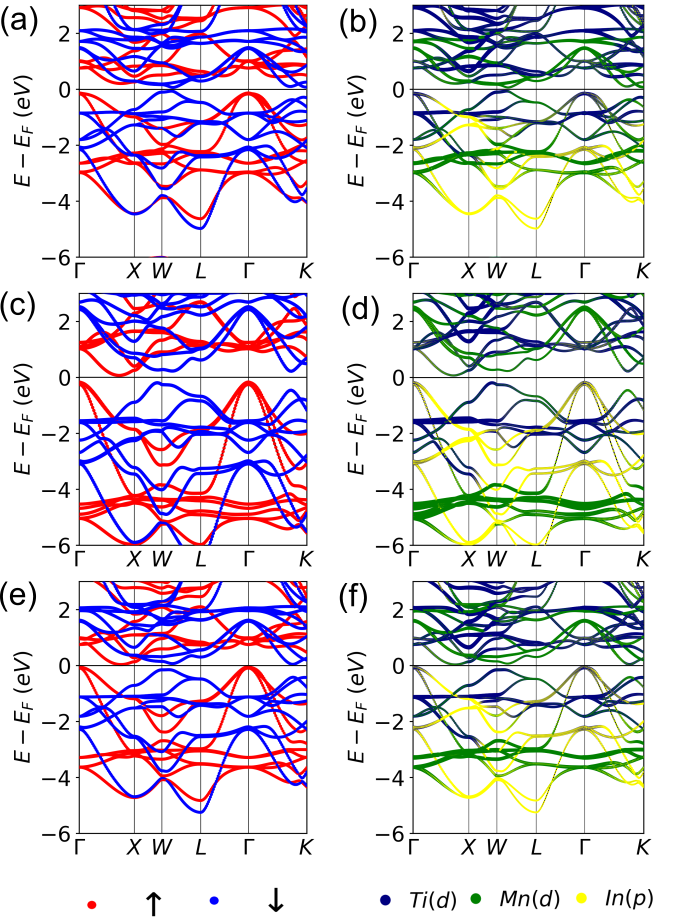}
    \caption{Ti$_2$MnIn band structures obtained with different DFT functionals:  PBE band structure projected on (a) the two spin channels and (b) atomic orbitals; HSE band structure projected on (c) the two spin channels and (d) atomic orbitals; PBE+U(BO) band structure projected on (e) the two spin channels and (f) atomic orbitals. Red and blue represent spin majority and minority, respectively.  Navy, green, and yellow represent Ti(d), Mn(d), and In(p) orbitals, respectively. 
    }
    \label{fig:Ti2MnIn_bands}
\end{figure}

\begin{figure}[h]
    \centering
    \includegraphics[width=0.45\textwidth]{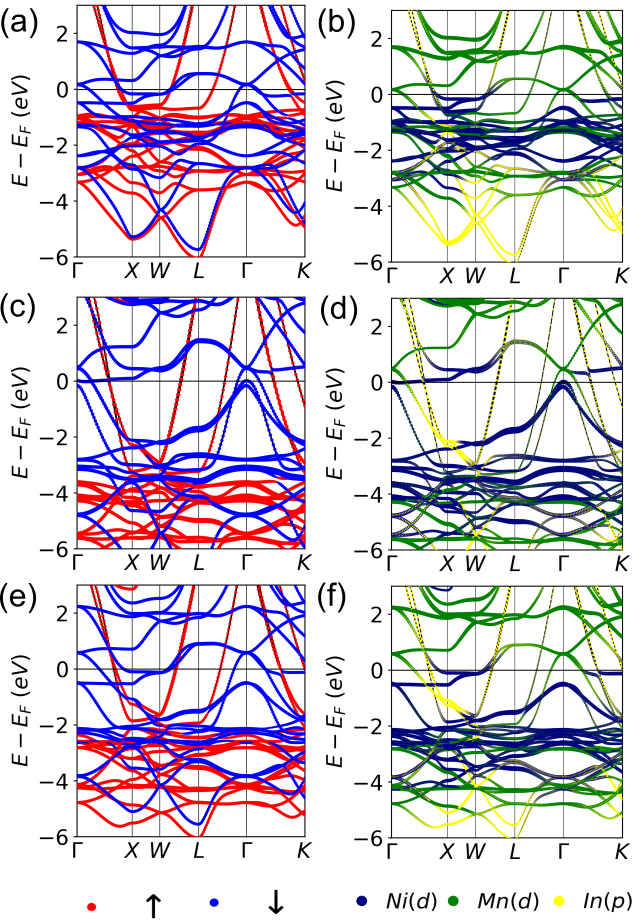}
    \caption{Ni$_2$MnIn band structures obtained with different DFT functionals:  PBE band structure projected on (a) the two spin channels and (b) atomic orbitals; HSE band structure projected on (c) the two spin channels and (d) atomic orbitals; PBE+U(BO) band structure projected on (e) the two spin channels and (f) atomic orbitals.  Red and blue represent spin majority and minority, respectively. Navy, green, and yellow represent Ni(d), Mn(d), and In(p) orbitals, respectively.}
    \label{fig:Ni2MnIn_bands}
\end{figure}

Figures \ref{fig:Ti2MnIn_bands} and \ref{fig:Ni2MnIn_bands} show the bulk band structures of Ti$_2$MnIn and Ni$_2$MnIn, obtained with different functionals. Density of states (DOS) plots are provided in Figures S2 and S3. Figure \ref{fig:Ti2MnIn_bands}a,b shows the band structure of Ti$_2$MnIn obtained with the PBE functional. In agreement with previous results \cite{Jia2014,Ma2018,Fang2014}, we find a ferrimagnetic material with a small overall magnetic moment of 0.04 $\mu_B$, and asymmetric gaps in the two spin channels.  We obtain a gap in the spin majority band of 0.339 eV, whereas the spin minority band has a gap of 0.128 eV (additional details are  are reported in Table S2).  These values differ slightly from the majority and minority gaps of 0.394 and 0.042 eV, respectively, reported in \cite{Ma2018}. Ref \cite{Ma2018} did not consider SOC, although we find that adding SOC to the PBE calculation has a minor effect on the gaps (see Table S2 in the SI).  Our two calculations differed slightly in the plane-wave energy cutoff, Brillouin zone sampling, and lattice constant, which explains the small difference in the results. 

Figure \ref{fig:Ti2MnIn_bands}c,d shows the band structure of Ti$_2$MnIn obtained with the HSE functional. We find notable differences between PBE and HSE. The HSE functional inverts the relative sizes of the gaps in the two spin channels, with the majority gap of 0.237 eV now smaller than the minority gap of 0.414 eV. This result suggests that Ti$_2$MnIn is not a nearly spin-gapless semiconductor, but rather a conventional semiconductor with a small gap. This decreases its relevance in the context of spin injection. Additional differences between PBE and HSE are found in the features of the band structure. The Mn and Ti $d$-bands in the HSE calculation are shifted farther away from the Fermi level compared to PBE, with the Mn(d) states affected more strongly by the addition of exact exchange. As a result of the shift in the position of the $d$ bands, the local valence band maxima change positions around the W and K points, and the conduction band minima change from near degenerate minima near W and K to a clear minimum along $\Gamma$-X.  In addition, the character of the valence band local maximum at the $\Gamma$ point changes from predominantly Ti(d) with PBE to predominantly In(p) with HSE and the character of the conduction band minimum changes from Mn(d) to Ti(d).  The magnetic moments on individual atoms are larger for HSE compared to PBE (see Supplementary Table S3), although the total magnetic moment of 0.085 $\mu_B$, obtained with HSE, remains relatively close to zero. 

Figure \ref{fig:Ti2MnIn_bands}e,f shows the band structure of Ti$_2$MnIn obtained with PBE+U(BO). The valence band maximum and conduction band minimum have the same locations as in the HSE calculation. The relative sizes of the gaps in the two spin channels resemble the HSE result, although the magnitudes of 0.071 eV in the majority channel and 0.192 eV in the minority channel are both smaller.  The Mn(d) and Ti(d) bands have an intermediary position between the PBE and HSE results relative to the Fermi level. In the valence band, the Ti(d) states roughly center around -0.8 eV, -1.5 eV, and -1.2 eV, respectively, with PBE, HSE, and PBE+U(BO). The Mn(d) states are centered around -2.5 eV, -4.5 eV, and -3.5 eV, respectively for PBE, HSE, and PBE+U(BO). The Ti(d) character of the conduction band minimum reproduces that of HSE. In contrast, at the local valence band maximum, the prominent Ti(d) character is similar to the PBE result.  The magnetic moments on the individual atoms also have intermediary values between HSE and PBE (see additional details in Table S3), although the overall moment of 0.015 $\mu_B$ is closer to zero than either. 

Figure \ref{fig:Ni2MnIn_bands}a,b shows the band structure of Ni$_2$MnIn obtained with the PBE functional.  In agreement with previous results, Ni$_2$MnIn is a metallic ferromagnet \cite{Kilian2000,Qawasmeh2012}.  The overall moment of 4.08 $\mu_B$ compares favorably with 3.91 $\mu_B$ reported in \cite{daSilva1988} using the von Barth-Hedin parameterization of the LDA without SOC \cite{VonBarth1972} and 4.16 $\mu_B$ reported in \cite{Kilian2000} using the Perdew-Zunger parameterization of the LDA  \cite{Perdew1981} with SOC. These values agree reasonably well with the experimental value of 4.34 $\mu_B$ \cite{Buschow1983}. The percentage of spin polarization  is extracted from the ratio between the minority and majority DOS at the Fermi level (see panels (a) and (b) of Figure S5 in the SI). We find that without SOC, the Fermi level spin polarization, extracted from the DOS, is 32\% in agreement with Ref. \cite{Qawasmeh2012}. SOC increases the Fermi level spin polarization to 71\%.

Figure \ref{fig:Ni2MnIn_bands}c,d shows the band structure of Ni$_2$MnIn obtained with the HSE functional.  Similar to the case of Ti$_2$MnIn, the most significant difference between the PBE and HSE band structures is the shift of the Ni(d) and Mn(d) bands away from the Fermi level in the HSE result. The band character in the minority (down) channel is generally preserved around the $\Gamma$ point and near the Fermi level, although the Ni(d) bands now cross the Fermi level in contrast to the Mn(d) bands in the PBE band structure. The HSE magnetic moment of 5.81 $\mu_B$ is substantially larger than both the PBE (4.08 $\mu_B$) and experimental (4.4 $\mu_B$) values. This is consistent with the tendency of hybrid functionals to favor high-spin states \cite{gao2016applicability, pokharel2022sensitivity}. The PBE band structure has several minority spin bands around the Fermi level, driving the polarization toward minority.  In contrast, in the HSE band structure many of the same Ni and Mn $d$ bands are pushed away from the Fermi level, inflecting the 30\% polarization at the Fermi level toward the majority spin channel.    

Figure \ref{fig:Ni2MnIn_bands}e,f shows the band structure of Ni$_2$MnIn obtained with PBE+U(BO). Similar to the case of Ti$_2$MnIn, the PBE+U(BO) result is intermediate between PBE and HSE. The Ni(d) bands are partially shifted toward the HSE result, although the character around the $\Gamma$ point and E$_f$ more closely resembles that of PBE. The Fermi level position is more centred between the Mn(d) and Ni(d) bands around $\Gamma$. The magnetic moment of 5.03 $\mu_B$ similarly has an intermediate value between PBE and HSE. The Fermi level spin polarization is 85\% minority, higher than both the PBE and HSE results. 

We next comment on the free electron spin injection analysis performed by Killian and Victora in \cite{Kilian2000}, and follow their model of momentum-accepting states in InAs. A full analysis is reported in Table S5 and the main results are summarized here. The transmission coefficient analysis is based on the band structure, therefore changes to the band structures resulting from the inclusion of SOC or from using different functionals indirectly affect the resulting transmission coefficients.  Our PBE results reproduce the transmission coefficients obtained in by Killian and Victora, although we observe an increase in T$_{[110]min}$ to 0.94, which we attribute to considering SOC.  In the  HSE band structure the Ni(d) bands are at the Fermi level, replacing the Mn(d) bands as those pertinent for this analysis.  The crossings are thus closer to the $\Gamma$ point, giving higher transmission coefficients of T$_{[100]min}$=0.96, T$_{[110]min}$=0.90, and T$_{[111]min}$=0.98, noting the final result is from the second crossing in that direction. In the PBE+U(BO) band structure, the Mn(d) bands cross the Fermi level, as in PBE, but the crossing is farther away from $\Gamma$.  As a result, the transmission coefficients are lower: T$_{[100]min}$=0.53, T$_{[110]min}$=0.71, and T$_{[111]min}$=0.95.  HSE and PBE+U(BO) generally drive the majority bands away from the Fermi level, stretching the downward-dispersing bands and pushing them closer to $\Gamma$.  As a consequence, these functionals produce an increased transmission in the majority spin channel. In particular, both the HSE and PBE+U(BO) yield transmission coefficients of 0.75 in the [111] direction.  These results diverge significantly from those reported in Ref. \cite{Kilian2000}, where the largest majority transmission coefficient was 0.39 in the [111] direction. The minority transmission remains robustly high in all channels even when a different band is sampled.  The majority transmission, however, changes qualitatively, depending on the DFT functional and may exceed 50\%. This has implications for the extent to which a spin polarized current could be injected from Ni$_2$MnIn into InAs. This further justifies the need for more detailed simulations of the interface.  

\subsection{\label{sec:level2}Interface model construction}

\begin{figure}[h]
    \centering
    \includegraphics[width=0.46\textwidth]{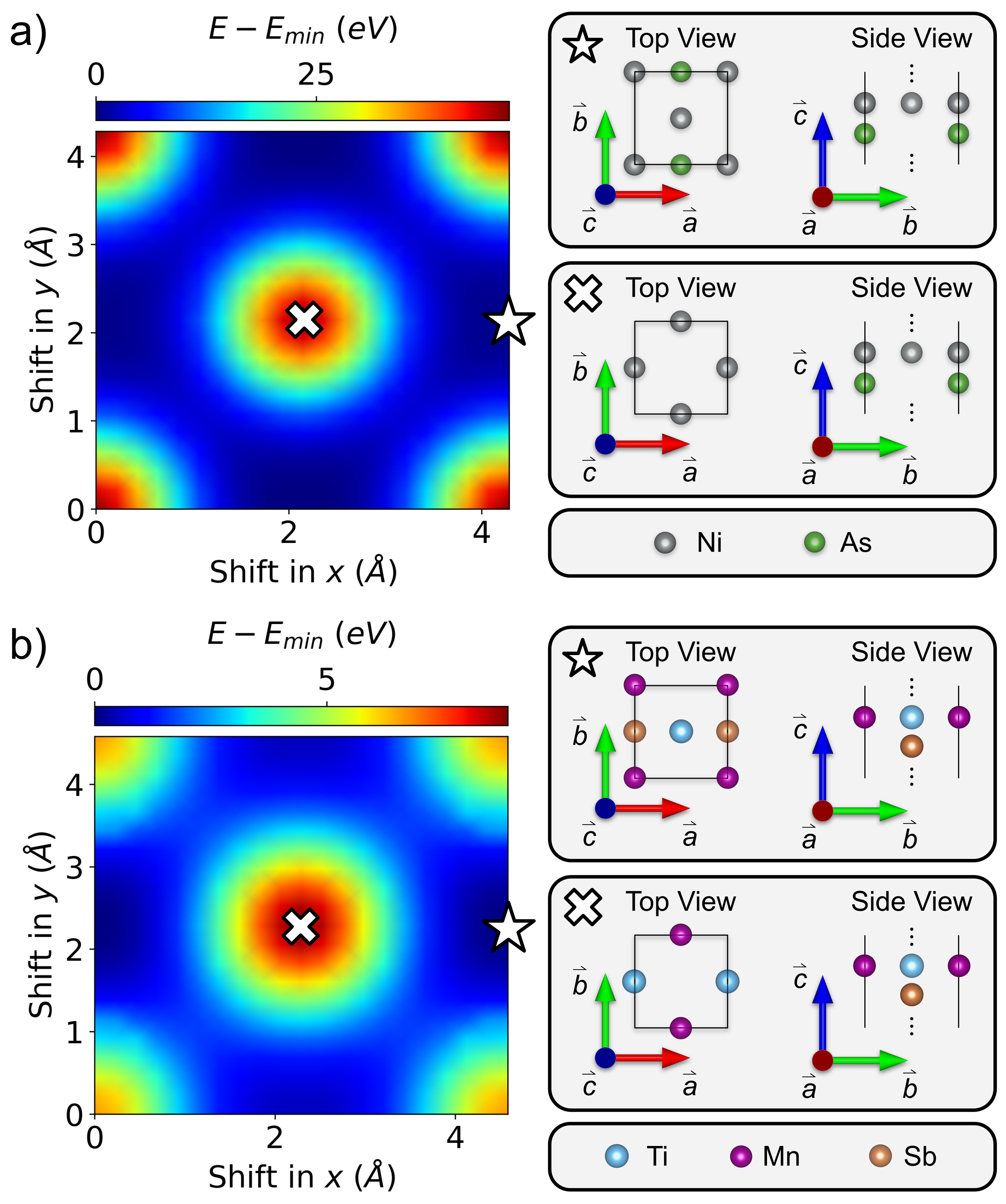}
    \caption{ Potential energy surface of (a) the Ni$_2$MnIn/InAs, Ni-As interface and (b) the Ti$_2$MnIn/InSb, TiMn-Sb interface, obtained by shifting the Heusler film on top of the III-V substrate in the $xy$ plane at the optimal interfacial distance.  The minimum, which corresponds to the most stable configuration, is marked by a star and the maximum is marked by an X. The corresponding structures are illustrated on the right.}
    \label{fig:surfaceopt}
\end{figure}

Slab models of the Ni$_2$MnIn/InAs and Ti$_2$MnIn/InSb interfaces were constructed with a (001) orientation. The interface models were constructed as periodic heterostructures with no vacuum region. The bulk lattice constants of 6.058 $\AA$ (ICSD 24518) for InAs and 6.48 $\AA$ for InSb (ICSD 24519) were used under the assumption that epitaxially grown Heusler films would conform to the substrate. The InAs and InSb substrates were As and Sb terminated, respectively. This is experimentally realizable with As(Sb)-rich growth conditions. For each interface, two terminations are possible for the Heusler.  Ni$_2$MnIn has alternating planes of Ni and MnIn along the [001] direction whereas Ti$_2$MnIn has alternating  planes of TiMn and TiIn. For both terminations of each material, the optimal interface distance was determined using dispersion-inclusive DFT. Curves of the energy as a function of the interfacial distance are shown in Figure S7. The minima for the Ni-As, MnIn-As, TiMn-Sb, and TiIn-Sb interfaces are obtained at 1.0, 1.8, 1.8, and 1.9 $\AA$, respectively. The optimal interface registry was found by translation of the Heusler slab with respect to the III-V semiconductor parallel to the interface for a model comprising six layers of semiconductor and four layers of Heusler.The results for the Ni$_2$MnIn/InAs, Ni-As interface and the Ti$_2$MnIn/InSb, TiMn-Sb interface are shown in Figure \ref{fig:surfaceopt}. Results for the other two interface terminations are provided in Figure S8. At the optimal configuration the Heusler atoms nestle in the pockets created by the group V atoms of the semiconductor. Finally, three layers of the Heusler and six layers of the semiconductor were relaxed on either side of the interface.

\begin{figure}[h]
    \centering
    \includegraphics[width=0.45\textwidth]{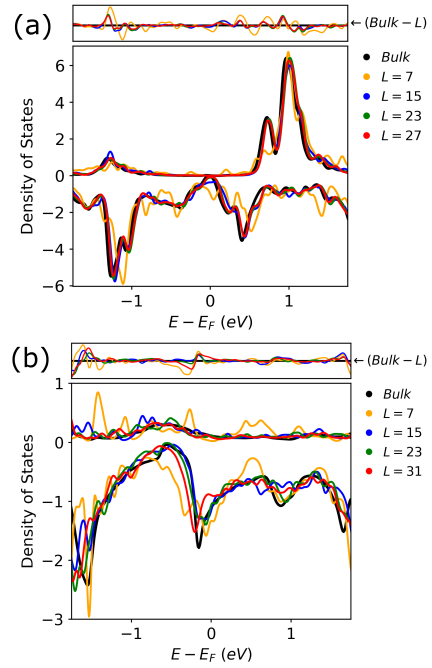}
    \caption{Density of states of the middle two layers of a surface slab model with $L$ atomic layers and a vacuum region of 60 $\AA$, compared to the bulk density of states for  (a) Ti$_2$MnIn using a 11x11x1 k-point grid and (b) Ni$_2$MnIn using a 9x9x1 k-point grid.}
    \label{fig:991_compare}
\end{figure}

To avoid size dependence due to the quantum size effect, once the interface configuration was optimized, the thickness of each material was increased until convergence of the electronic properties to their bulk values was achieved. Once the minimal thickness required to achieve convergence is reached, any further changes to the electronic structure with increasing thickness are insignificant. 47 atomic layers of the semiconductor were used, based on previous band gap convergence studies \cite{Yang2021a,Yang2021,YuMarom2021a}. The  Heusler thickness was converged with respect to the density of states and magnetic moment, which should approach the bulk limit at the center of the slab \cite{Dardzinski2022}.  For Ti$_2$MnIn, the convergence test was performed with the TiIn termination. We note that the surface termination should not affect the convergence of the DOS and magnetic moment at the center of the slab. To avoid spurious states in the band gap of Ti$_2$MnIn, dangling bonds on the surface were passivated by pseudo-hydrogen atoms \cite{Huang2005}. Charges of 1.0 $e$ and 1.25 $e$ were used to terminate the surface Ti and In atoms, respectively. A k-point grid of 11x11x1 was used.  As shown in Figure \ref{fig:991_compare}a, the density of states in the middle of the Ti$_2$MnIn slab reaches the bulk limit by 23 layers. For the interface model reported below, 25 layers are used.  The density of states for Ni$_2$MnIn, terminated with Ni, as a function of the number of layers is shown in Figure \ref{fig:991_compare}b for a 9x9x1 k-point mesh.  The DOS for 23 and 31 layers are closer to the bulk limit than those for 7 and 15 layers.  The interface model reported below comprises 25 layers. The DOS obtained from a similar calculation using a 21x21x1 k-point mesh, shown in the SI,  is closer to the bulk DOS. However, owing to the high computational cost of the interface slab calculation with a total of 99 atoms, a 9x9x1 k-point mesh is used below. For both materials, the magnetic moments approach the bulk values with increasing thickness, as shown in Figure S9. The number of atomic layers, 47 for the semiconductor and 25 for the Heusler, was chosen such that the construct was symmetric under a quarter rotation about the layer in the middle of each material. This ensured that the two interfaces in each periodic heterostructure were identical.

\begin{figure*}
    \centering
    \includegraphics[width=0.8\textwidth]{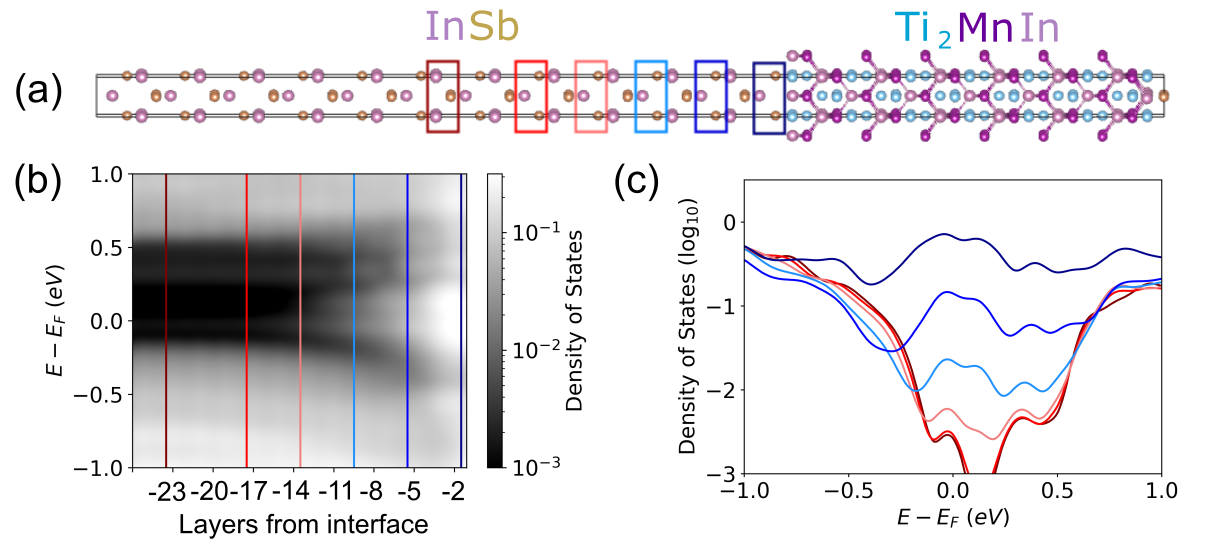}
    \caption{
    Electronic structure of the Ti$_2$MnIn/InSb TiIn-Sb interface. (a) Illustration of the interface model.  (b) Density of states in the InSb as a function of atomic layer, where the layers are numbered based on distance from the interface, which is located at zero.  (c) Local density of states for selected layers, indicated by boxes and lines in the same colors in panels (a) and (b), respectively. 
        }
    \label{fig:TMI-InSb_TiIn_DOS}
\end{figure*}

\begin{figure*}
    \centering
    \includegraphics[width=0.8\textwidth]{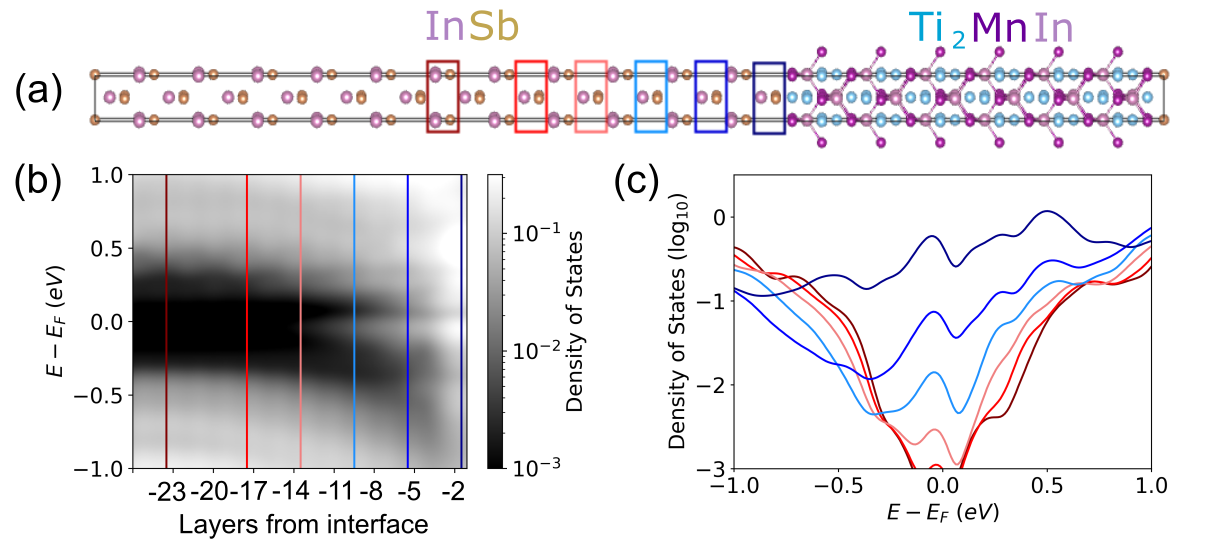}
    \caption{
    Electronic structure of the Ti$_2$MnIn/InSb TiMn-Sb interface. (a) Illustration of the interface model.  (b) Density of states in the InSb as a function of atomic layer, where the layers are numbered based on distance from the interface, which is located at zero.  (c) Local density of states for selected layers, indicated by boxes and lines in the same colors in panels (a) and (b), respectively.  
    }
    \label{fig:TMI-InSb_TiMn_DOS}
\end{figure*}

\subsection{\label{sec:TMI}Ti$_2$MnIn/InSb}

Figures \ref{fig:TMI-InSb_TiIn_DOS} and \ref{fig:TMI-InSb_TiMn_DOS} show the density of states as a function of distance from the interface for the InSb side of the Ti$_2$MnIn/InSb TiIn-Sb interface and TiMn-Sb interface, respectively. Similar plots for the Ti$_2$MnIn side of the interface are provided in Figures S10-S11. For both interface terminations, the InSb does not exhibit significant band bending at the interface. For the TiIn-Sb interface, the band gap of the InSb is above the Fermi level in the middle of the construct, whereas for the TiMn-Sb interface the Fermi level is in the middle of the InSb gap. Bader analysis \cite{Sanville2007} shows a charge of 0.028 $e$ per supercell transferred from Ti$_2$MnIn to InSb for the TiIn-Sb interface and 0.019 $e$ per supercell transferred from Ti$_2$MnIn to InSb for TiMn-Sb interface. Although this charge transfer is expected to raise the Fermi level in InSb, comparison to the Ni$_2$MnIn/InAs interface below, shows the charge transfer is relatively small, thus not significantly affecting the Fermi level position. Although the Ti$_2$MnIn is not metallic, because its band gap is smaller than that of InSb, states from the Ti$_2$MnIn penetrate into the InSb, similar to metal induced gap states (MIGS). These states decay with the distance from the interface and completely vanish by layers 17-18.     

\begin{figure}[h]
    \centering
    \includegraphics[width=0.4\textwidth]{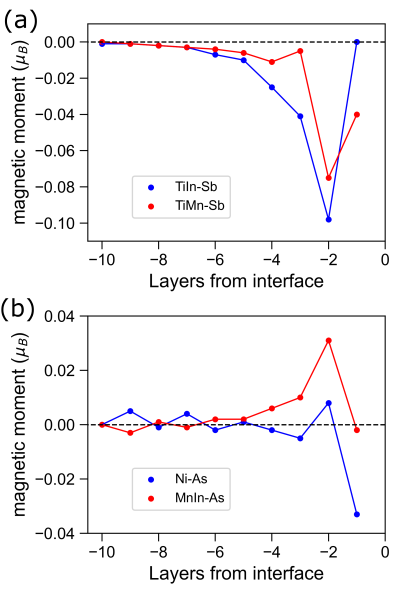}
    \caption{The induced magnetic moments in the III-V semiconductor as function of number of atomic layers from the interface:  (a) InSb in Ti$_2$MnIn/InSb and (b) InAs in Ni$_2$MnIn/InAs. }
    \label{fig:moment_plots}
\end{figure}

Figure \ref{fig:moment_plots}a shows the magnetic moment in the InSb as a function of distance from the interface.  Weak spin polarization is induced in the first few layers of InSb.  The proximity induced magnetic moments for the TiIn interface of 0.000 $\mu_B$ on the first Sb layer and -0.098 $\mu_B$ on the second In layer from the interface are smaller by orders of magnitude than the magnetic moments of -1.693 $\mu_B$ on Ti in the first layer of Ti$_2$MnIn (the In moment in the first layer of Ti$_2$MnIn is -0.116 $\mu_B$).  Similarly, for the TiMn interface, the induced moments in the semiconductor are -0.040 (-0.075) $\mu_B$ for Sb (In), compared to the magnetic moments in the Heusler of -1.632 (3.627) $\mu_B$ for Ti (Mn).  Magnetic moment values across the interface are tabulated in Tables S6-S9.  The interface configuration does not have a significant effect on the magnetic moment induced in the semiconductor, as the moments in the TiIn (TiMn) layers of Ti$_2$MnIn are ferromagnetically (antiferromagnetically) coupled and thus have a similar net moment. The induced magnetic moments found here are of the same order of magnitude reported previously for the EuS/InAs interface \cite{YuMarom2021a} and the Fe/InSb interface \cite{Moayedpour2021}. 

\begin{figure}[h]
    \centering
    \includegraphics[width=0.45\textwidth]{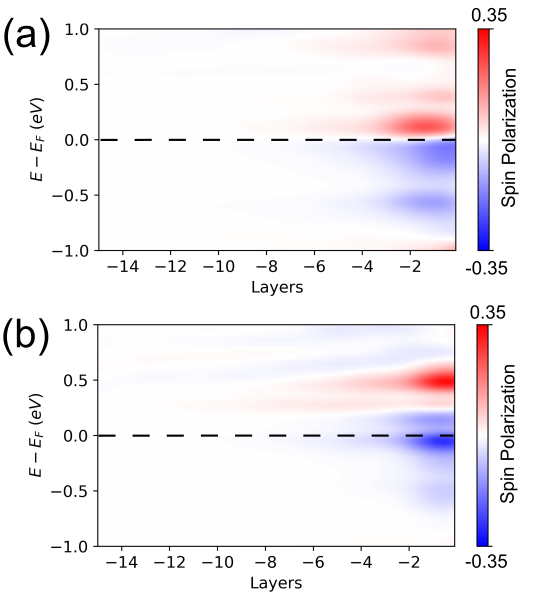}
    \caption[TMI-InSb spin polarization]
    { Spin polarization, defined as the difference between the majority DOS and minority DOS [n($\uparrow$) - n($\downarrow$)], as a function of the distance from the interface for the InSb side of the Ti$_2$MnIn/InSb interface for (a) the TiIn termination and (b) the TiMn termination.     }
    \label{fig:TMI-InSb_spin_pol}
\end{figure}

Spin polarized transport is important for spintronics. In Ref. \cite{Schultz2009FeGaAs} the computed spin polarization at the Fermi level was correlated with the experimentally observed spin polarization of electrons injected from Fe into GaAs through an interface phase of Fe$_3$Ga. Figure \ref{fig:TMI-InSb_spin_pol} shows the spin polarization, \textit{i.e.,} the difference between the majority DOS and minority DOS, as a function of the distance from the interface for the InSb side of the Ti$_2$MnIn/InSb interface. For this paper, "majority" spin in the semiconductor refers to spins with the same orientation as those in the majority channel in the Heusler compound.  For both the TiIn and TiMn terminations spin polarization is induced in the InSb in the vicinity of the interface, vanishing by around six layers into the semiconductor. The spin polarized region spatially overlaps with the region in which states from the Ti$_2$MnIn penetrate into the InSb (see Figures \ref{fig:TMI-InSb_TiIn_DOS} and \ref{fig:TMI-InSb_TiMn_DOS}). Owing to the localized nature of the proximity induced magnetism, differences are observed between the two interface terminations. The TiIn interface exhibits majority spin polarization above the Fermi level and minority spin polarization below the Fermi level.  The TiMn interface exhibits minority spin polarization around the Fermi level.  The spin polarization around the Fermi level in Ti$_2$MnIn is spin down, as shown in both the primitive cell DOS and the layer-resolved spin polarization plots in Figure S16.  Based on this, we surmise that minority spin transport, possibly with high transmission, may occur across the TiIn-Sb interface. For the TiMn-Sb interface, it may be possible apply a bias to tune into a highly transmissive minority transport regime or into a majority transport regime with the interface acting as a spin filter, similar to the case of the Fe/GaAs interface, as discussed in \cite{Schultz2009FeGaAs}. We note that both interface terminations may be present in a real system, which could make it impossible to resolve the differences in their transport behavior.

\begin{figure*}
    \centering
    \includegraphics[width=0.8\textwidth]{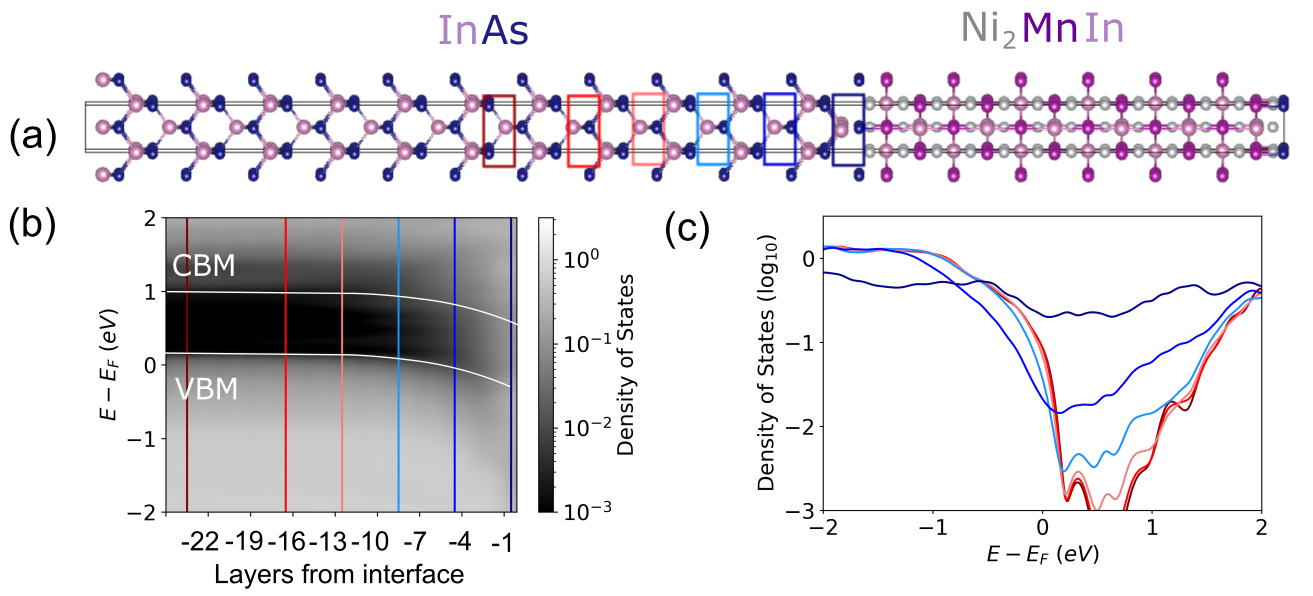}
    \caption{
    Electronic structure of the Ni$_2$MnIn/InAs Ni-As interface. (a) Illustration of the interface model.  (b) Density of states in the InAs as a function of atomic layer, where the layers are numbered based on distance from the interface, which is located at zero.  (c) Local density of states for selected layers, indicated by boxes and lines in the same colors in panels (a) and (b), respectively. The lines indicating the band bending are a guide to the eye.
    }
    \label{fig:NMI-InAs_NiAs_DOS}
\end{figure*}

\begin{figure*}
    \centering
    \includegraphics[width=0.8\textwidth]{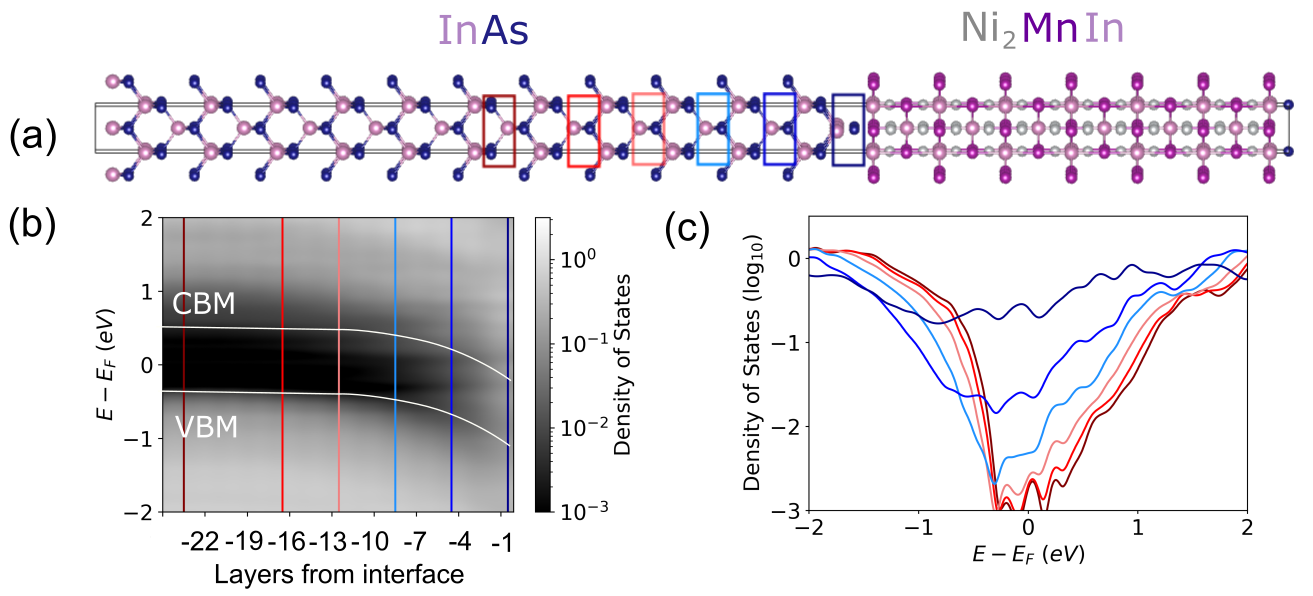}
    \caption{
    Electronic structure of the Ni$_2$MnIn/InAs MnIn-As interface. (a) Illustration of the interface model.  (b) Density of states in the InAs as a function of atomic layer, where the layers are numbered based on distance from the interface, which is located at zero.  (c) Local density of states for selected layers, indicated by boxes and lines in the same colors in panels (a) and (b), respectively. The lines indicating the band bending are a guide to the eye.
    }
    \label{fig:NMI-InAs_MnIn_DOS}
\end{figure*}

\subsection{\label{sec:NMI}Ni$_2$MnIn/InAs}

Figures \ref{fig:NMI-InAs_NiAs_DOS} and \ref{fig:NMI-InAs_MnIn_DOS} show the density of states as a function of distance from the interface for the InAs side of the Ni$_2$MnIn/InAs interface for the Ni-As interface and the MnIn-As interface, respectively. As expected, both interfaces exhibit band bending in the InAs \cite{Brillson1986} with the Fermi level shifting toward the conduction band in the vicinity of the interface.   Similar to the Ti$_2$MnIn/InSb interface, the Heusler termination affects the Fermi level position and the charge transfer across the interface. The Fermi level lies in the valence band deep in the InAs when interfaced with Ni, whereas it is in the middle of the band gap when interfaced with MnIn.  Bader analysis shows a charge of 0.16 $e$ per supercell transferred from Ni$_2$MnIn to InAs for the MnIn interface and a charge of 0.243 $e$ per supercell transferred from InAs to Ni$_2$MnIn for the Ni-As interface.  The latter indicates a depletion of charge in the semiconductor and lowering of the Fermi level. MIGS penetrate from the Ni$_2$MnIn into the InAs. These decay with the distance from the interface and completely vanish by layers 17-18.

\begin{figure}[h]
    \centering
    \includegraphics[width=0.45\textwidth]{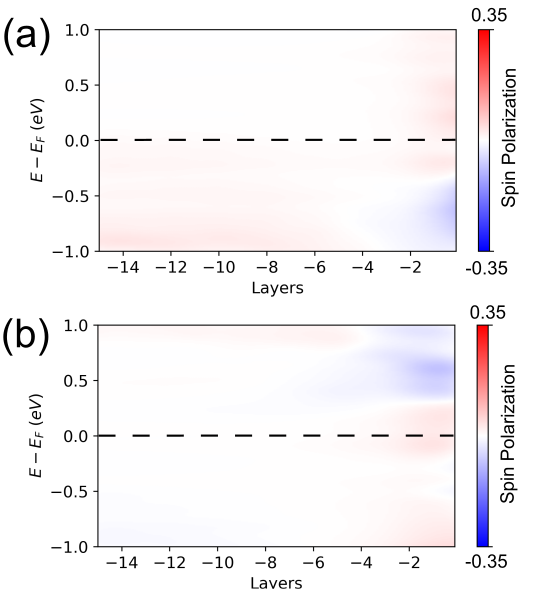}
    \caption[NMI-InAs spin polarization]
    {Spin polarization, defined as the difference between the majority DOS and minority DOS [n($\uparrow$) - n($\downarrow$)], as a function of the distance from the interface for the InAs side of the Ni$_2$MnIn/InAs interface for (a) the NiAs termination and (b) the MnIn termination.     }
    \label{fig:NMI-InAs_spin_pol}
\end{figure}

The induced magnetic moments in the InAs as a function of the distance from the interface are shown in Figure \ref{fig:moment_plots}b. Tabulated values are provided in Tables S6-S9. More significant differences between the two terminations of the Heusler are found for the Ni$_2$MnIn/InAs interface than for the Ti$_2$MnIn/InSb interface. When adjacent to Ni, the first layer of As from the interface holds a moment of -0.033 $\mu_B$. For the MnIn interface,  the magnetic moment of -0.002 $\mu_B$ on the first As layer is close to zero.  The situation is reversed for the second In layer from the interface.  For the MnIn interface the second In layer has a magnetic moment of 0.031 $\mu_B$, whereas the In layer closest to the Ni interface has a nearly zero moment of 0.008 $\mu_B$. The significant differences between the two terminations may be explained by the relatively high magnetic moment of about 4 $\mu_B$ on the Mn atoms in the Heusler compared to the much smaller magnetic moments of 0.024 $\mu_B$ and -0.114 $\mu_B$ on the first layer of Ni.  

Figure \ref{fig:NMI-InAs_spin_pol} shows the spin polarization , \textit{i.e.,} the difference between the majority DOS and minority DOS, as a function of the distance from the interface for the InAs side of the Ni$_2$MnIn/InAs interface.  Both interface terminations exhibit a slight spin polarization at the Fermi level next to the interface, which decays within about five layers, or a couple of nanometers. The spin polarized region overlaps with with the location of the MIGS.  For both Ni$_2$MnIn terminations majority spin polarization is found around the Fermi level in the InAs.   The spin polarization for the Heusler side of this interface, provided in Figure S16, shows minority spin polarization around the Fermi level, which is opposite to that found in the InAs.  In correspondence with the discussion of the Fe/GaAs interface in \cite{Schultz2009FeGaAs}, this interface may act as a spin filter, rotating the spins as the interface is traversed. Surprisingly, for both Ni$_2$MnIn terminations, the magnetic behavior induced in the first few layers of InAs is weaker than in the InSb, both in terms of the magnetic moment and the spin polarization, despite the ferromagnetic nature of Ni$_2$MnIn compared to the ferrimagnetic Ti$_2$MnIn. The induced magnetic moments on atoms may be explained by the highly localized nature of the magnetic interactions at the interface. The difference in the induced spin polarization around the Fermi level may be explained by the fact that despite being overall ferromagnetic, the Ni$_2$MnIn is weakly spin polarized around the Fermi level (see Figure S16) because the Mn $d$ states are relatively far from the Fermi level (see Figure \ref{fig:Ni2MnIn_bands}), whereas some of the Mn $d$ states in Ti$_2$MnIn are in the vicinity of the Fermi level. This demonstrates the importance of considering the atomistic details of the interface, not just the bulk properties of the isolated materials.

\section{\label{sec:conclusion}Conclusion}

We have presented a first-principles study of epitaxial Heusler/III-V interfaces of Ti$_2$MnIn/InSb and Ni$_2$MnIn/InAs using DFT with a machine-learned Hubbard U correction determined by Bayesian optimization. For the bulk materials, adding the Hubbard U correction improves upon the performance of the semilocal PBE functional and produces band structures closer to the HSE hybrid functional. The main effect of the addition of exact exchange in the hybrid functional and of the Hubbard U correction is a shift of the transition metal $d$ bands away from the Fermi level. This has qualitative implications for the electronic structure of Ti$_2$MnIn. With PBE it has a near zero gap in the minority spin channel, however with HSE and PBE+U(BO) it is a conventional narrow-gap semiconductor. Ni$_2$MnIn remains a ferromagnetic metal but HSE and PBE+U(BO) change the character and curvature of the states around the Fermi level, which may have implications for transport through the interface. Importantly, the computational cost of PBE+U(BO) enables us to study large interface models, which would not be feasible with a hybrid functional.

Both the Ti$_2$MnIn/InSb and Ni$_2$MnIn/InAs interfaces exhibit states penetrating from the Heusler into the gap of the semiconductor, which decay within 18 atomic layers. As expected from branching point theory \cite{tersoff1986calculation, 1996_Monch_JApplPhys}, the InAs shows significant band bending whereas the InSb does not. Both interfaces exhibit weak and localized proximity induced magnetism in the semiconductor. The magnetic moments induced on the atoms in the second layer of the semiconductor from the interface are less than 0.1 $\mu_B$. This is of the same order of magnitude found previously for the for the EuS/InAs interface \cite{YuMarom2021a} and the Fe/InSb interface \cite{Moayedpour2021}. For both interfaces, spin polarization, \textit{i.e.}, a difference between the majority and minority DOS, is induced around the Fermi level of the semiconductor. The spin polarized region spatially overlaps with the MIGS and decays within a few atomic layers. Counter-intuitively, although Ni$_2$MnIn is ferromagnetic and Ti$_2$MnIn is ferrimagnetic, the magnetism induced in the first few layers of InAs is weaker than in the InSb, both in terms of the magnetic moment and the spin polarization. This may be explained by the fact that, despite being overall ferromagnetic, the Ni$_2$MnIn is weakly spin polarized around the Fermi level  because the Mn $d$ states are relatively far from the Fermi level. In comparison, some of the Mn $d$ states in Ti$_2$MnIn are in the vicinity of the Fermi level, producing a more significant spin polarization in that energy range. 

For the Ni$_2$MnIn/InAs interface, the InAs is slightly spin polarized at the Fermi level with majority polarization, opposite to the minority polarization around the Fermi level in the Ni$_2$MnIn. This means that the interface may act as a spin polarizer, similar to the Fe/GaAs interface \cite{Schultz2009FeGaAs}. For the Ti$_2$MnIn/InSb interface the sign of the induced spin polarization depends on the termination of the Heusler. For the TiMn termination, minority spin polarity is induced in the InSb, which is the same as the polarization of the Ti$_2$MnIn around the Fermi level. This means the interface may be highly transmissive. For the TiIn termination the induced polarity is majority above the Fermi level and minority below the Fermi level. As a result, it may be possible to switch the spin polarization of a current going through the interface by applying a voltage bias. We note that it is possible that both terminations would be present simultaneously in a realistic system, depending on whether the termination can be controlled by modifying the growth conditions. Our results demonstrate that owing to the localized nature of the magnetic interactions at the interface, it is important to consider the atomistic details of the structure and the electronic structure around the Fermi level, rather than the overall properties of the bulk materials.   

In the context of Majorana devices, the criteria described in the introduction are not met. The few nanometer range of the induced magnetism in the semiconductor is smaller by two orders of magnitude than the typical InAs and InSb nanowire diameter of around 100 nm. We interpret this as signifying that the weak and local magnetic interactions at the interface are likely insufficient to drive proximity-induced magnetism over the body of the wire. These results are in agreement with previous work on EuS/InAs \cite{YuMarom2021a}.  Moreover, even if the relevant states to produce a Majorana zero mode were confined to the region of the wire that is close to the interface, the presence of MIGS, which render the semiconductor gapless, may preclude tuning the wire into a single subband limit \cite{Frolov2020}. Therefore, we conclude that both the Ti$_2$MnIn/InSb and Ni$_2$MnIn/InAs interfaces are of limited usefulness in the context of Majorana-based quantum computing devices. Nevertheless, these interfaces and other Heusler/III-V interfaces may be of interest for spintronics because they offer the possibility of manipulating spin currents through "band structure engineering" by controlling the local structure and composition of the interface. In particular, our findings demonstrate that hitherto overlooked ferrimagnetic Heusler materials may still be capable of inducing spin polarization in a semiconductor, depending on the extent of spin polarization at the Fermi level. This calls for further computational and experimental exploration of Heusler/semiconductor interfaces.

\section{\label{sec:level5} Acknowledgments}

 We thank Chris Palmstr{\o}m from the University of California Santa Barbara for helpful discussions. This research was funded by the Department of Energy through grant DE-SC0019274. This research used resources of the National Energy Research Scientific Computing Center (NERSC), a DOE Office of Science User Facility supported by the Office of Science of the U.S. Department of Energy under contract no. DE-AC02-05CH11231.

\bibliography{manuscript}

\end{document}